\documentclass[fleqn,usenatbib]{mnras}

\usepackage{newtxtext,newtxmath}

\usepackage[T1]{fontenc}
\usepackage{ae,aecompl}

\usepackage{graphicx}	
\usepackage{amsmath}	
\usepackage{amssymb}	

\newcommand{\lat}{{\textit {Fermi}}}

\newcommand{\snr}{SNR\,G39.2--0.3}
\newcommand{\fermi}{4FGL\,J1903.8+0531}

\title[SNR G39.2--0.3 in gamma-rays]{SNR G39.2--0.3, an hadronic cosmic rays accelerator}

\author[E. de O\~na Wilhelmi et al.]{
Emma de O\~na Wilhelmi$^{1,2,3}$\thanks{E-mail: emma.de.ona.wilhelmi@desy.de},
Iurii Sushch$^{3,4,5}$\thanks{E-mail: iurii.sushch@desy.de},
Robert Brose$^{3,6}$,
Enrique Mestre$^{1,2}$,
\newauthor
Yang Su$^{7}$
and
Roberta Zanin$^{8}$
\\
$^{1}$Institute of Space Sciences (ICE/CSIC), Campus UAB, Carrer de Can Magrans s/n, 08193 Barcelona,
Spain\\
$^{2}$Institut d'Estudis Espacials de Catalunya (IEEC), 08034 Barcelona, Spain \\ 
$^{3}$Deutsches Elektronen Synchrotron DESY, 15738 Zeuthen, Germany\\
$^{4}$Centre for Space Research, North-West University, 2520 Potchefstroom, South Africa\\
$^{5}$Astronomical Observatory of Ivan Franko National University of Lviv, Kyryla i Methodia 8, 79005 Lviv, Ukraine\\
$^{6}$Institute of Physics and Astronomy, University of Potsdam, 14476 Potsdam, Germany\\
$^{7}$Purple Mountain Observatory and Key Laboratory of Radio Astronomy, Chinese Academy of
Sciences, Nanjing 210034, China\\
$^{8}$CTA Observatory GmbH, Via Piero Gobetti 93, I-40129 Bologna, Italy}
\date{Accepted XXX. Received YYY; in original form ZZZ}

\pubyear{2020}

\begin{document}
\label{firstpage}
\pagerange{\pageref{firstpage}--\pageref{lastpage}}
\maketitle

\begin{abstract}
Recent results obtained with gamma-ray satellites have established supernova remnants as accelerators of GeV hadronic cosmic rays. In such processes, CRs accelerated in SNR shocks interact with particles from gas clouds in their surrounding. In particular, the rich medium in which core-collapse SNRs explode provides a large target density to boost hadronic gamma-rays. \snr\ is one of the brightest SNR in infrared wavelengths, and its broad multiwavelength coverage allows a detailed modelling of its radiation from radio to high energies. We reanalysed the \emph{Fermi}-LAT data on this region and compare it with new radio observations from the MWISP survey. The modelling of the spectral energy distribution from radio to GeV energies favours a hadronic origin of the gamma-ray emission and constrains the SNR magnetic field to be at least $\sim100$ $\mu$G. Despite the large magnetic field, the present acceleration of protons seems to be limited to $\sim10$~GeV, which points to a drastic slow down of the shock velocity due to the dense wall traced by the CO observations, surrounding the remnant. Further investigation of the gamma-ray spectral shape points to a dynamically old remnant subjected to severe escape of CRs and a decrease of acceleration efficiency. The low-energy peak of the gamma-ray spectrum also suggests that that the composition of accelerated particles might be enriched by heavy nuclei which is certainly expected for a core-collapse SNR. Alternatively, the contribution of the compressed pre-existing Galactic cosmic rays is discussed, which is, however, found to not likely be the dominant process for gamma-ray production.

\end{abstract}

\begin{keywords}
ISM: individual objects: G39.2--0.3, ISM: supernova remnants, gamma-rays: ISM, clouds: ISM
\end{keywords}



\section{Introduction}
\label{sec:intro}
The deep survey performed by the $Fermi$ LAT telescope has revealed a large population of gamma-ray bright SNRs \citep{2016ApJS..224....8A}. From these gamma-ray observations, two trends on the SNR population were established, based on their gamma-ray spectra and physical characteristics of the remnant: a population of young, hard-spectrum SNRs, and a second population of older and brighter, often interacting with dense molecular clouds (MC), ones. These evolved SNRs have lost much of their energy budget, but the dense surrounding media enhance the gamma-ray emission via proton-proton interaction.  The spectral energy distribution (SED) from this emission process is characterized by a sharp rise in the $\sim$70--200 MeV range (resulting from the neutral pion energy threshold production), followed by a hard emission up to the maximum energy, which is determined by either the maximum energy to which the CRs are accelerated or by the escape of high energy protons into the interstellar medium. The detailed spectral investigation of a handful of those SNR-MC systems have resulted on the detection of such feature, strongly favouring hadronic processes as the origin of the gamma-ray emission observed \citep{2013Sci...339..807A,2011MmSAI..82..747G,2016ApJ...816..100J,2019A&A...623A..86A}. A second possibility to explain the observed gamma-ray spectrum is via bremsstrahlung radiation, in which electrons instead of protons interact with the surrounding media. However, to explain the sharp drop observed, an artificial lower cut-off in the spectrum of electrons has to be invoked \citep{2019A&A...623A..86A}. The emission mechanisms that dominate  SNRs depend on many different aspects such as the shock evolution and the circumstellar medium. The fast shock velocity promotes in principle the acceleration of CRs, both of hadronic or leptonic nature. Whether the gamma-ray emission is originated by proton-proton interactions or by electrons off-scattering soft photon fields by inverse Compton mechanisms, or via bremsstrahlung, is still subject of discussion for the brightest sources observed. In general, the global spectral shape found in bright and young SNRs seems to favour a leptonic origin. 
Still, a hadronic interpretation cannot be completely ruled out under certain conditions \citep{2011ApJ...734...28A,2019MNRAS.487.3199C}. The spectra of other young but much fainter remnants, such as Cassiopeia A or Tycho, seem to be better explained by hadronic interaction \citep{2017MNRAS.472.2956A,2017ApJ...836...23A,2014ApJ...785..130Z,2012A&A...538A..81M}.

\snr\ belongs to a group of core-collapse SNRs, classified as Type IIL/b SNe \citep{2009ApJ...691.1042L}. Type IIL/b SNe go through a phase of red supergiant (RSG), creating a rich and highly in-homogeneous medium in which the remnant expands. Together with SNR G11.2--0.2, RCW 103, W44 and W49B, this SNR shows a bright NIR H$_2$ emission, evidencing the presence of shocked molecular gas \citep{2009ApJ...691.1042L,2011ApJ...727...43S,1990A&A...232..467P,1993ApJ...408..514A}. Based on X-ray and radio mm observations, the age of SNR was estimated to be in the 3 to 7 kyr old range \citep{2011ApJ...727...43S}. 
There is a large uncertainty about its distance, which was estimated using different X-ray and radio observations \citep{1975A&A....45..239C,2003ApJ...592L..45O,1975AJ.....80..679B,1997AJ....114.2058G}. We adopted thus a reference distance of 6.2 kpc.

\begin{figure*}
  \centering
  \includegraphics[width=\linewidth]{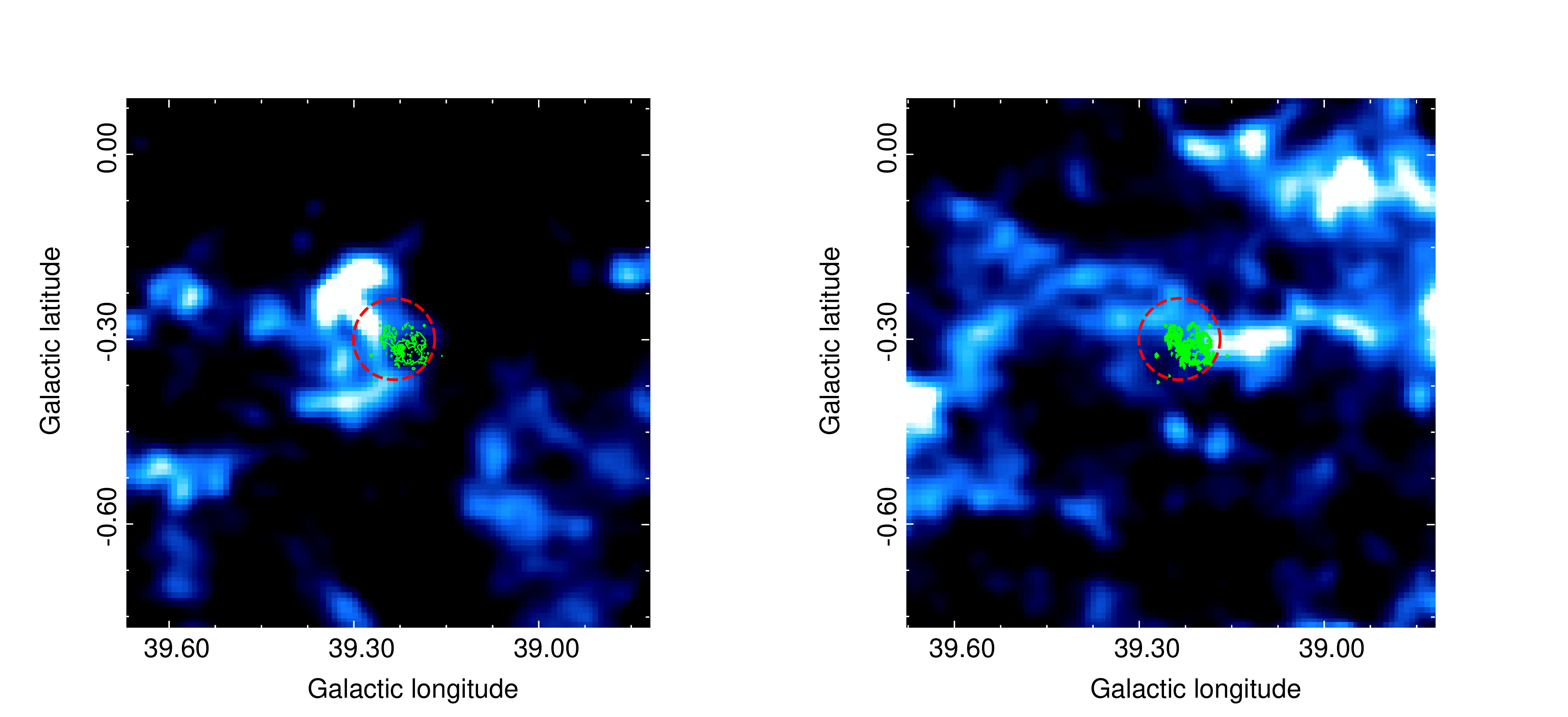}
   \includegraphics[width=\linewidth]{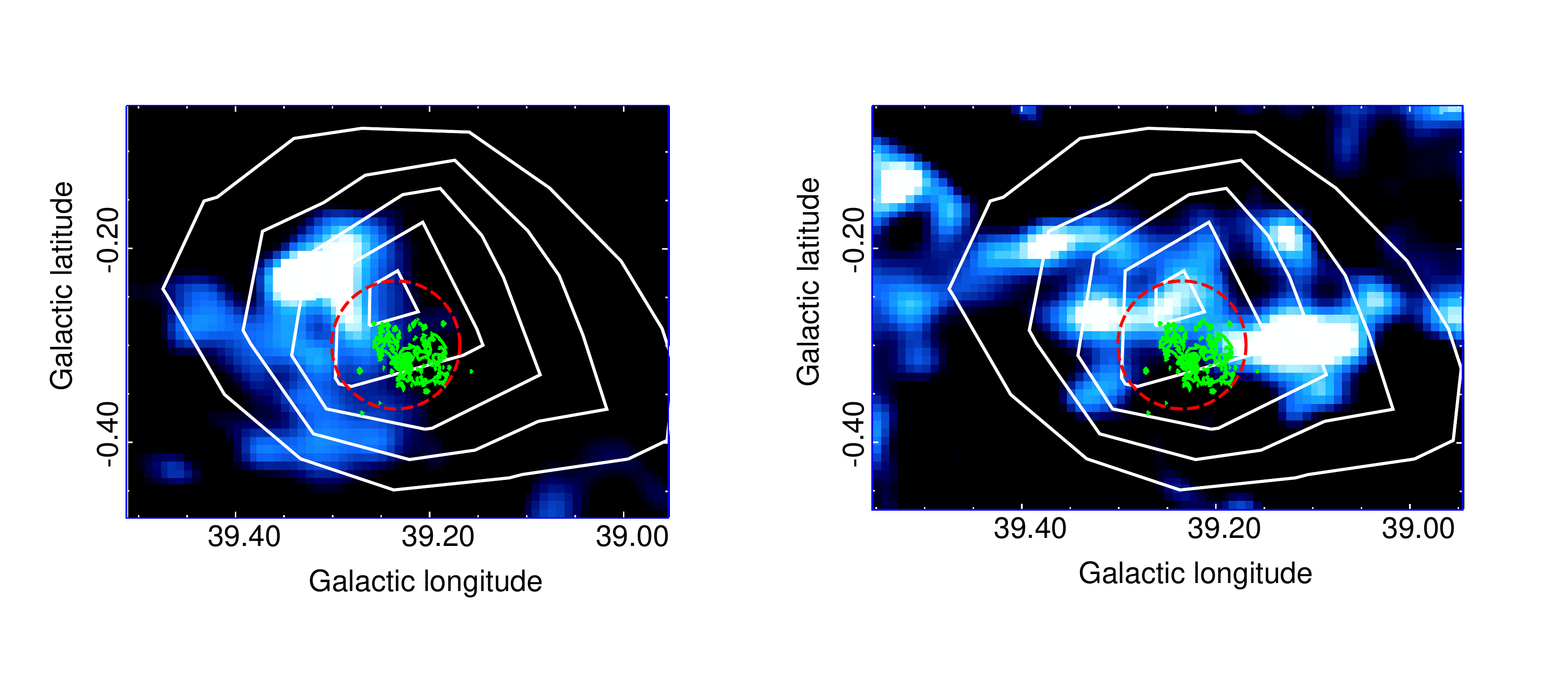}
  \caption{On the top left (a) $^{13}$CO large-scale intensity map integrated between 67 km s$^{-1}$ and 74 km s$^{-1}$ obtained from the MWISP survey. On the top right (b) the intensity map integrated between 80 and 88 km s$^{-1}$. The red dashed-line circle marks the 99 per cent localization error of the reanalysis of \fermi. The green contours correspond to the radio shell, as derived from the image compiled by \citealt{2012AdSpR..49.1313F}. On the bottom we showed the $^{12}$CO channel (MWISP survey), centred on the position of \snr. The left one (c) for the 69 km s$^{-1}$ map and the right one (d) for the 88 km s$^{-1}$ one. White contours are obtained from LAT significance map above 3\,GeV, starting on TS=25 (or $\sim5\sigma$ in steps of 10).}
 \label{fig:CO}
\end{figure*}

The large and extensive bibliography on \snr\ \footnote{http://www.mrao.cam.ac.uk/surveys/snrs/snrs.G39.2-0.3.html} provides a broad range of information. It is an asymmetric composite SNR with its central part dominated by non-thermal emission from the likely associated pulsar wind nebula \citep{2003ApJ...592L..45O, 2020MNRAS.492.1484S}. The western shell is brighter in radio and X-ray, suggesting a molecular cloud interaction, whereas it shows a faint tail on the eastern side. The remnant has been largely observed using molecular tracers. \citealt{2009ApJ...691.1042L} remarked a cavity-like structure at V$_{\rm {LSR}} \sim$69 km s$^{-1}$ surrounding the remnant. The location of the cavity suggests an associated multishell structure produced by the interactions of the SNR shocks with the RSG wind material. This interpretation was refuted by \citealt{2011ApJ...727...43S}, who proposed a different association with a region located at V$_{\rm {LSR}}\sim$88 km s$^{-1}$. At this somehow larger distance, the western boundary seems to be  confined by a molecular wall, whereas the fading material to the east matches the X-ray and radio image. \citealt{1999ApJ...516..811H} estimated a supernova explosion energy of $E\sim3.5\times10^{50}$ ergs, for a filling factor of 1 ($E\sim7\times10^{50}$ ergs for the filling factor of 0.25 expected for Sedov-Taylor solutions), and equipartition magnetic field of $B\sim100$ $\mu$G ($B\sim140$ $\mu$G for the filling factor of 0.25), based on the study of the observed X-ray thermal emission. These estimates are however very uncertain as they rely on the assumption of the pure Sedov-Taylor solution and uniform density. Interaction with a massive molecular cloud would certainly increase the estimates of the explosion energy and equipartition magnetic field by increasing the density of the medium and thus decreasing the filling factor of the X-ray emitting plasma inside the remnant. 
\cite{2016MNRAS.459.4224C} studied the microwave emission from the remnant and described the spectral energy distribution with a two components, a Synchrotron spectrum with index $\alpha=0.36$ (compatible with previous radio measurements) and a thermal blackbody component for dust temperature of 25\,K and optical depth at 100$\mu$m of 3$\cdot$10$^{-4}$.  The bright emission from \snr\ is also listed in the Spitzer survey of SNRs in our Galaxy \citep{2006AJ....131.1479R} and results on a FIR density of u$_{\rm FIR}$ of $\sim$2.4~eV/cm$^3$.  

In the high energy regime, the \lat\ 4FGL catalogue listed a source located within 0.1$^{\rm o}$ of the SNR, \fermi. The gamma-ray spectrum is represented with a {\it LogParabola} function \citep{2019arXiv190210045T} and it appears as likely associated with \snr. No detection has been claimed at higher energies in the TeV regime \citep{2018A&A...612A...3H}, where an upper limit of 6$\cdot$10$^{-13}$ erg/cm$^2$s was derived from H.E.S.S. observations of the Galactic plane. Recently, \citealt{2020MNRAS.492.1484S} reanalysed the region, obtaining similar spectral parameters to the one listed in the 4FGL catalogue.

In the following we explore the association between the high-energy source and \snr\ and investigate the physics mechanisms powering the gamma-rays observed.  

\section{Data Analysis and Results}
\label{sec:data}
\subsection{High energy emission}
\label{sec:lat}
We analyzed $\sim$11 yr of data spanning from 2008 August 4 2008 to 2019 June 19 with energies between 100\,MeV and 100\,GeV. The dataset was analysed using \textsc{Fermipy}\footnote{http://fermipy.readthedocs.io/en/latest/} v0.17.3: a set of python programmed tools that automatize the PASS8 analysis with the Fermi Science Tools\footnote{https://fermi.gsfc.nasa.gov/ssc/data/analysis/documentation/}.  FRONT+BACK events class was used to increase the statistics. We used P8R3\_SOURCE\_V2 instrument response function. We applied the energy dispersion to all sources in our model expect for the isotropic (extragalactic) diffuse model. To study the morphology, we selected events with energy above 3 GeV to optimize the instrument angular resolution\footnote{https://www.slac.stanford.edu/exp/glast/groups/canda/lat\_Performance.htm}. Once the best-fitting position and extension is determined, we selected a larger energy range, from 100 MeV to 100 GeV to obtain the spectrum. The latest was derived by performing a maximum likelihood analysis in a circular region (15$^{\rm o}$) around the best-fitting position. The emission model for our radius of interest includes the LAT sources listed in the fourth LAT catalogue (4FGL, \citealt{2019arXiv190210045T}) within a region of 30\degr\ radius around \snr\ and the diffuse $\gamma$-ray background models; the Galactic diffuse emission modelled by \textit{gll\_iem\_v07.fits} 
and the isotropic component by \textit{iso\_P8R3\_SOURCE\_V2\_v1.txt}, including the instrumental background and the extragalactic radiation. 

We used the tools \emph{localize} and \emph{extension}, which perform a scan of the likelihood surface around the nominal position and compute the likelihood ratio test with respect to the no-extension (point-source) hypothesis, respectively. To obtain the spectral energy distribution, we split the total energy range into 14 logarithmically spaced bins. We required that each spectral point has at least a Test Statistics TS$\,= 4$, otherwise a 95 per cent confidence level (CL) upper limit was computed. 

The study of the location and extension results on a point-like source, centred at RA$_{\rm J2000} = (286.01  \pm  0.02)^{\rm o}$ and Dec$_{\rm J1000} = (5.47 \pm 0.02)^{\rm o}$ with an error in the localization at 99 per cent of 0.066 (see red dashed line in Fig. \ref{fig:CO}). The 99 per cent confidence circle contains the \snr\, strongly suggesting a physical association between the SNR and the gamma-ray source. 

\begin{figure}
  \centering
  \includegraphics[width=8.0cm]{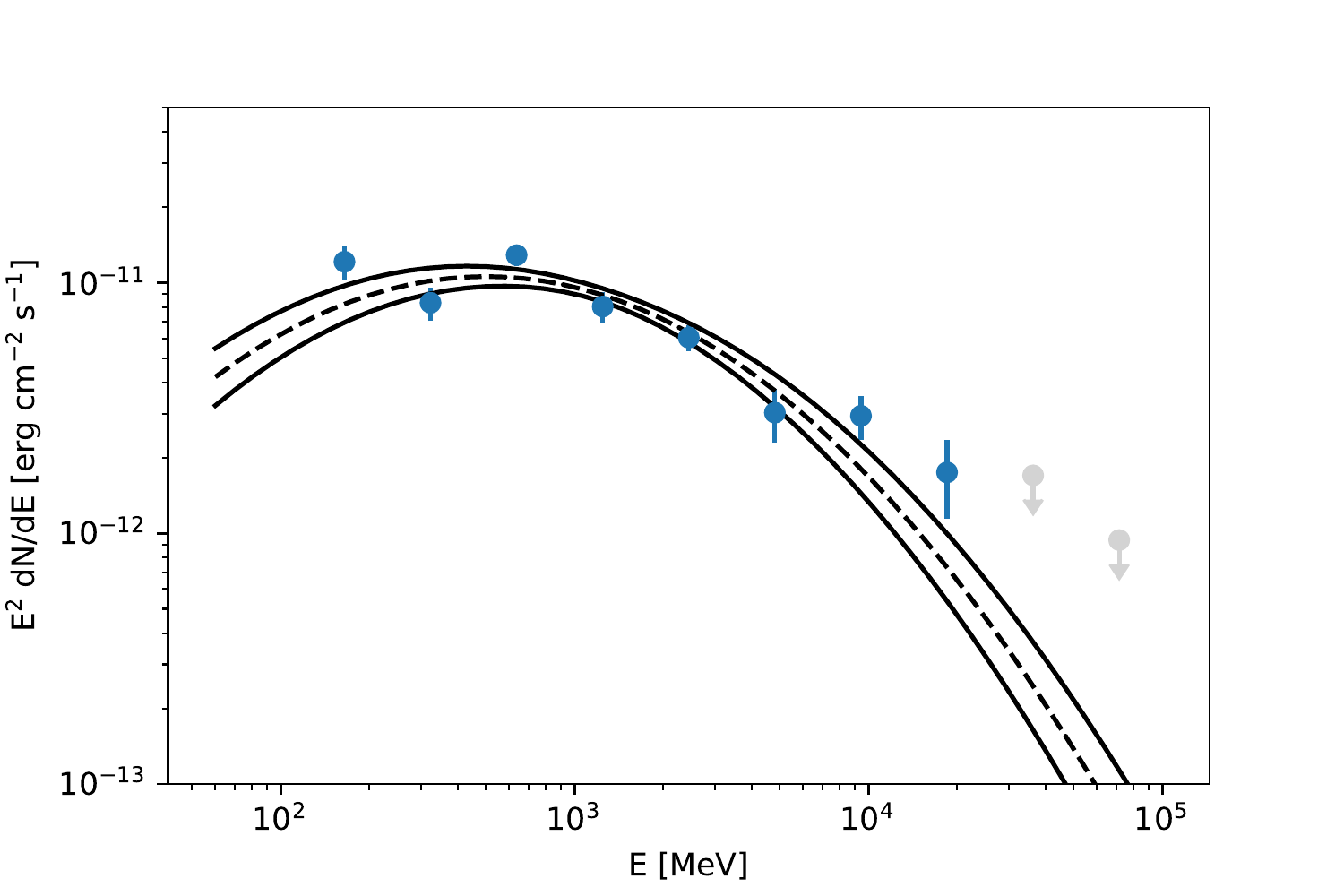}
  \caption{Spectral energy distribution of \fermi, associated to \snr. The best-fitting shape and the 1$\sigma$ error derived for the energy range between 100 MeV and 100 GeV is represented by black lines.}
  \label{fig:spectrum}
\end{figure}

Fig. \ref{fig:spectrum} shows the spectral energy distribution obtained with the likelihood method and using the above-described morphology as extraction regions. The differential energy spectrum is well represented by a LogParabola function such $dN/dE = N_{\rm o}$(E/E$_{\rm break})^{-(\alpha + \beta \rm{log}(E/E_{\rm break}}$), where the best-fitting values are: N$_{\rm o}=(8.0\pm0.7)\cdot10^{-13}$ MeV$^{-1}$/cm$^2$s, $\alpha = 2.6 \pm$0.1 and $\beta =0.20 \pm$0.03, for a break energy value of E$_{\rm break}=2.3$\,GeV. The source is detected with a TS of 363 above 100 MeV.  The fit values are in good agreements with the ones found by \citealt{2020MNRAS.492.1484S} and the ones obtained from the 4FLG catalogue \citep{2019arXiv190210045T}.

\subsection{CO observations}
\label{sec:co}  
The large-scale CO map is obtained from the MWISP survey (see details in \citealt{2019ApJS..240....9S}). The $^{12}$CO(J=1--0), $^{13}$CO(J=1--0), and C18O (J=1--0) lines were simultaneously observed using the 13.7\,m telescope. The covered region is towards the northern Galactic plane with $|b|< 5^{\rm o}$. The half-power beam width (HPBW) is $\sim$50 arcsec at the frequency of $\sim110-115\,\rm{GHz}$. The rms noise is $\sim0.5\,$K for $^{12}$CO and $\sim$0.3\,K for $^{13}$CO and C18O, at a velocity resolution of $\sim$0.2\,km s$^{-1}$ with a uniform grid spacing of 30 arcsec. 

To evaluate the dense medium in which the SNR is located, we derived the skymap around its position integrating the 3D cube in the two velocity ranges previously proposed: between 67 and 74 km s$^{-1}$ (as discussed in \citealt{2009ApJ...691.1042L}), and between 80 and 88 km s$^{-1}$ (as proposed by \citealt{2011ApJ...727...43S}). 
To obtain the mass density from the two intensity maps, we use the standard assumption of a linear relationship between the velocity-integrated CO intensity, W$_{\rm CO}$, and the column density of molecular hydrogen, N(H2), adopting for the conversion factor X$_{\rm CO}  = 2.0 \cdot 10^{20}$  cm$^2$ (K km s$^{-1}$) \citep{2001ApJ...547..792D,2013ARA&A..51..207B}. Given the uncertainties in the distance, we used a fiducial value of 6.2\,kpc for both velocity ranges, which translates on a physical size of $\sim$16 pc (for a 0.15$^{\rm o}$ source). The integrated mass density of the two velocity ranges is similar, with 440 cm$^{-3}$ and 326 cm$^{-3}$ for the 69 km s$^{-1}$ and 88 km s$^{-1}$ range, respectively. 

To understand the giant molecular cloud (GMC) structure around the SNR and the high-energy source, we used the $^{13}$CO(J=1--0) channel. Given the complicated velocity field in the Galactic plane, the optical thin channel is thus a better tracer of the different molecular clouds with different velocities (that is, different kinematic distance) at large scales.  The two large-scale intensity maps for the 69 km s$^{-1}$ and 88 km s$^{-1}$ velocity ranges are showed in Fig. \ref{fig:CO}(a) and (b) respectively. The two GMC proposed are clearly visible in the two maps, being the first one (a) slightly offset ($\sim0.15^{\rm o}$) to the North with respect to the position of the \lat\ source and \snr. For a detailed view of the SNR surroundings, we used the $^{12}$CO(J=1--0) channel, that traces the denser region (Fig \ref{fig:CO} (c) and (d), for the 69 km s$^{-1}$ and 88 km s$^{-1}$ velocity ranges respectively). The gas distribution seems to be more scattered along \fermi\ favouring an association with this second region, based on the gas and gamma-ray similar distribution.

\section{Spectral Modelling}
\label{sec:naima}
\begin{figure}
  \centering
  \includegraphics[width=8.0cm]{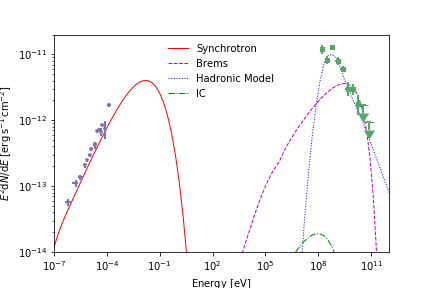}
  \caption{Multiwavelength spectral energy distribution of the \snr. The lines indicate the best-fitting models for both leptonic and hadronic interpretation of the high energy emission.}
  \label{fig:spectrumMWL}
\end{figure}

The optimal agreement between the gamma-ray source and \snr\ points to a clear association of the two sources. Moreover, the dense region in which the SNR is embedded, suggests a scenario in which the SNR is accelerating CRs, electrons or hadrons, that interact with the rich medium around the remnant. The high energy spectrum is characterised with a quite narrow photon distribution (see Fig. \ref{fig:spectrum} and Fig. \ref{fig:spectrumMWL}, peaking at a few hundreds of MeV. To investigate the nature of the parent particles, we fit the obtained gamma-ray spectral energy distribution of the SNR by both hadronic- and leptonic-induced gamma-ray spectrum. We use the {\it Naima} package (version v0.8.3, \citealt{2017ascl.soft08022Z}) to compute the non-thermal spectral distribution from radio to TeV energies. The X-ray non-thermal emission detected by Chandra is not shown, since it is likely correlated to the associated PWN and not with the shell SNR. The spectral shape of the LAT source makes very unlikely a PWN origin, under the assumption of a Synchrotron/inverse Compton model (see green dashed line in Fig. \ref{fig:spectrumMWL} for the inverse Compton contribution). An unrealistic (few hundreds of eV/cm$^3$ instead of the $\sim$2 eV/cm$^3$ measured) photon field density in the FIR region should be up-scattered to boost the inverse Compton flux to the level of the gamma-ray flux observed. The shape of the low-energy part of the electron spectrum is actually well determined by the fit to the radio data, obtained from the shell region. We used a power-law function with spectral index fixed to 1.8 (corresponding to a Synchrotron spectrum of 0.4) and an exponential cut-off to account for the maximum energy of the accelerated particles within the shell. The corresponding bremsstrahlung emission cannot reproduce the LAT spectrum at high energies, but the latter constrains the amplitude of the electron population to $<3\cdot10^{34}$ eV$^{-1}$ for a density of $\sim$400 cm$^{-3}$ (obtained from the CO observations) in the cloud, witch is believed to be interacting  with the remnant. The combined radio and GeV fit also limits the magnetic field in the SNR to at least $B \ge 150 \mu G$ to reach the level of Synchrotron emission observed for the population of electrons. Even if assuming a 50 per cent error in the estimation of the molecular content, a large magnetic field above 100$\mu G$ is still required to reach the radio emission level. This value is comparable with the one estimated by equipartition arguments \citep{1999ApJ...516..811H} assuming a constant in time density of the ambient medium of $0.5$~cm$^{-3}$ extracted from observations of thermal X-rays. This value of magnetic field is larger than what one would expect for an evolved SNR, but still not that surprising for an SNR interacting with a dense cloud.
?still within the values expected for SNRs of a few thousands years old in a dense region

The magnetic field strengths inside molecular clouds deduced from Zeeman observations by \cite{2010ApJ...725..466C} indicates roughly constant magnetic fields of $B_0=10\,\mu$G in clouds with densities of $n_0\lesssim 300\,\text{cm}^{-3}$ (similar to  the density of the cloud interacting with \snr). This field can be further compressed by a factor of a few by the interaction with the SNR shock. If the shock is already radiative, a dense shell will form in the downstream of the shock, where the field can be further compressed. Assuming a turbulent field in the cloud yields $B_{\rm d}=B_0\sqrt{\frac{2\xi^2+1}{3}}$, where $\xi$ is the density-compression ratio between the cloud and the radiative shell. For the not unreasonable case of $\xi=15$, one can reach the field strengths of $\approx 150\,\mu$G needed to explain the Synchrotron emission. Another possible explanation for a high magnetic field strength is the amplification of a turbulent field by MHD instabilities in the downstream of the shock propagating in an inhomogeneous medium \citep{2014NIMPA.742..169F, 2007ApJ...663L..41G, 2019MNRAS.487.3199C}. However, the fields provided by these processes should only convert a fraction of the kinetic (and thermal) energy density of the downstream flow into magnetic fields. 
In general one expects
\begin{equation}
    B_{\rm d} \approx \sqrt{\frac{4\pi n m_p \sigma}{\xi}}v_\text{shock},
\end{equation}%
where $\sigma$ is the conversion efficiency. A shock velocity of $\approx150\,$km s$^{-1}$ and a conversion efficiency of $\sigma$=5 per cent yields $B_{\rm d}\approx150\,\mu$G for $\xi=4$ and $B_{\rm d}\approx80\,\mu$G for $\xi=15$.
It is not clear if the compression of the cloud's field or the amplification of the field by MHD-instabilities will dominate the contribution to the magnetic field. However, the combination of both processes ensures a reasonably high magnetic field to explain the observed Synchrotron emission.

Similarly, the combination of the radio and GeV data constrains the maximum electron energy to $E_{\rm cut} \le 100\,$GeV (assuming the electron spectrum follows a power law with an exponential cut-off). Such a high magnetic field would imply severe Synchrotron losses which should modify the electron spectrum introducing a break, but the anticipated break energy would be roughly similar to the maximum energy constrained above and hence taking Synchrotron losses into account would not radically change the picture.

Contrary to the leptonic models, we obtain a satisfactory fit of the high energy data when using hadronic-originated ones. The best fit obtained is plot in Fig. \ref{fig:spectrumMWL} in blue dotted line. For the injected population of protons, we use a broken power law. Such a spectral shape would be naturally expected as a result of a superposition of different cut-off energies during the SNR evolution and subsequent particle escape at later times, or by intrinsic properties of the CR acceleration in the shock (i.e. finite size of the emission region or evolution of the shock in partially ionized medium) \citep{2019MNRAS.490.4317C,2011MNRAS.410.1577O,1996A&A...309..917A,2019ApJ...874...50Z,2020A&A...634A..59B,2011NatCo...2..194M}. The total energy, normalised to a distance of 6.2 kpc, stored in protons (for E$_{p}>100$\,MeV) results  on W$_{\rm p}=3.2^{+1.1}_{-0.8}10^{49}$ erg, which represents only a few percent of the total energy of the SNR explosion. The best-fit for the two spectral indices are $s_1=2.0\pm1.5$ and $s_2=2.78\pm0.06$, for a break energy at E$_{\rm b}=220\pm70$\,MeV. The low break energy results on a large uncertainty on $s_1$, which is poorly constraint. For simplicity, we also fit the LAT spectrum with a power-law function, resulting on an equally good fit with a spectra index of $s=2.75^{+0.04}_{-0.06}$ and similar particle energy above 100\,MeV as the broken power-law case. This arises as a consequence of the low-energy cut-off in the gamma-ray spectrum from pion decays, which makes $s_1$, the spectral index obtained in particle acceleration process, irrelevant for the mathematical fit.

The acceleration of protons and electrons should occur in a similar way, implying a similar spectral shape (at least at lower energies at which electrons are not affected by Synchrotron cooling). This also implies that the spectral break derived for the proton spectrum should also be reflected in the electron spectrum at similar energies. The observed radio spectrum shows a featureless power-law spectrum up to 33 GHz \citep{2016MNRAS.459.4224C} constraining the energy of a possible break in the electron spectrum. The characteristic energy of the emitted Synchrotron photon is
\begin{equation}
    \nu_{\rm ch} \simeq 0.8 \left(\frac{E}{1\,\mathrm{GeV}}\right)^2 \left(\frac{B}{150\,\mu\mathrm{G}}\right)~\mathrm{GHz},
\end{equation}
implying that the spectral break cannot occur below $\sim 6 \left(\frac{B}{150\,\mu\mathrm{G}}\right)^{-1/2}$~GeV. Hence, to secure a break at a sub-GeV or even a few GeV level one would need to dramatically increase the magnetic field. The plausibility of such a break in the particle spectrum is further discussed in the next section in more detail.

\section{Discussion}
\label{sec:discussion}

Given the large magnetic field derived from the comparison between the radio and high energy emission, the low energy cut-off measured makes, in principle, \snr\ an extremely inefficient in accelerating particles into very-high energies. For a type II SNR of 3 to 7 kyr, exploding in a circumstellar wind of density $n(R) \propto R^{-2}$, shock velocities of the order of $10^3$ km s$^{-1}$ could still be expected (see i.e. \citealt{2016arXiv161007638G} and references therein). This large shock velocity together with the amplified magnetic field should be enough to accelerate protons to energies of more than TeV. On the contrary, the low cut-off measured suggests either a slow shock, or a large escape rate of TeV protons in the surrounding. This effect could be due to the evolution of the SNR in the dense media of its progenitor. Indeed, the comparison between the gamma-ray emission and the intensity maps obtained in CO (in Fig. \ref{fig:CO}) revealed a clear enhancement of molecular material on the position of the gamma-ray source. The point-like morphology of the gamma-ray emission, centred on the SNR, instead on the peak of the radio intensity observed on the 67 to 74 km s$^{-1}$ map (Fig. \ref{fig:CO} (a) and (c)), suggests an association with the second velocity range proposed (between 80 and 88 km s$^{-1}$, Fig. \ref{fig:CO}  (b) and (d)) corresponding to a distance of 6.2 kpc. The multi-wavelength investigation carried on by \citealt{2011ApJ...727...43S} revealed a thick molecular wall at this velocity, coincident with the bright X-ray, IR and radio emission from the west part of the remnant. Such thick wall could naturally explain a hampering of the SNR expansion, which would limit its acceleration power to very high energies and also the large magnetic field needed to account for the radio emission. 

In the following, we compare our results with different scenarios that account for the spectral shape of the derived hadronic component.

\subsection{Old dynamical age scenario}
\label{sec:old}
The shape of the observed gamma-ray spectrum with a turn-over at low energies ($E_{\mathrm{break}} = 2.3$~GeV in the LogParabola fit) suggests a high dynamical age of the remnant, i.e. the SNR is on the post-adiabatic stage of its evolution featuring a low shock velocity and substantial escape of CRs. Indeed, the slow shock results in a low maximum energy of freshly accelerated protons, while high-energy particles which are already in the system escape to far upstream of the shock. Recent detailed modeling of the time evolution of the CR spectrum in SNRs in post-adiabatic phase which benefited from a self-consistent treatment of the diffusion coefficient by solving the transport equation for magnetic turbulence induced by Alfvenic waves \citep{2020A&A...634A..59B}, showed that inefficient confinement of high-energy particles at late stages of evolution leads to the rapid decrease of the maximum energy reachable in the shock acceleration and hence to the formation of a spectral break at $10-100$~GeV. Above the break the spectrum softens and can be adequately approximated by a power law with a spectral index of about $2.7$. The break energy corresponds to the maximum energy of particles reachable in the acceleration process at the current stage. This spectral structure is similar to the one observed for SNR G39.2$-$0.3, where the observed gamma-ray spectrum can be explained by hadronic emission from a particle spectrum with the spectral index of $s_2=2.78\pm0.06$ above the break. The fit of the data, however, suggests an extremely low break energy of $E_{\rm b}=220\pm70$\,MeV, implying that particles are no longer being effectively accelerated. This estimate of the break energy is rather uncertain given that the gamma-ray spectrum can be equally well fitted with a simple power-law particle spectrum with the spectral index $s = 2.75^{+0.04}_{-0.06}$.

To examine this farther, we use the post-processing radiation routine of the \textsc{RATPaC} code \citep{2012APh....35..300T, 2013A&A...552A.102T, 2016A&A...593A..20B, 2018A&A...618A.155S} designed for numerical simulations of particle acceleration in SNRs. The module to calculate gamma-ray radiation from pion decays relies on  Monte Carlo event generators, namely {\sc DPMJET}-III \citep{roesler2001monte} and UrQMD \citep{1998PrPNP..41..255B, bleicher1999relativistic}, for the calculation of inelastic cross sections and differential production rates of secondary particles produced in nuclei collisions \citep{2019ICRC...36..592P}. We use this module to calculate the expected gamma-ray emission from a toy proton spectrum which follows a broken power law in momentum,

\begin{equation}
    \frac{dN}{dp} = 
    \begin{cases}
        N_0 p^{-s_1},& \text{if } p< p_{\rm b}\\
        N_0 p_{\rm b}^{-s_1+s_2} p^{-s_2},& \text{otherwise},
    \end{cases}
    \label{eq:particle_spectrum}
\end{equation}
where $p_{\rm b}$ is the the break momentum of a particle with the kinetic energy $E_{\rm b}$.

\begin{figure}
  \centering
  \includegraphics[width=8.0cm]{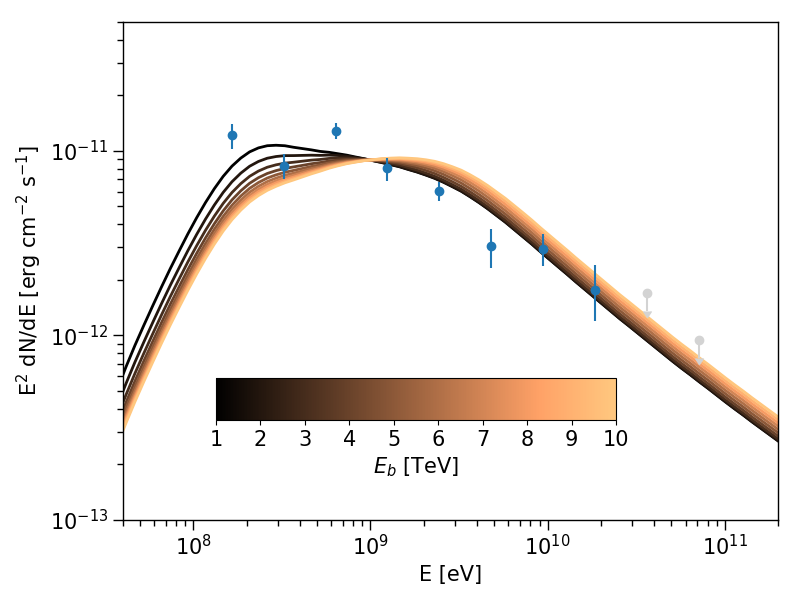}
  \caption{Spectral energy distribution of the modelled hadronic gamma-ray emission generated by a proton population described by a broken power-law spectrum in momentum with spectral indices $s_1 = 2.0$ and $s_2 = 2.75$. Different colours of lines denote different assumed values of the break energy from 1~GeV to 10~GeV. Blue points represent $Fermi$-LAT data analysed in this paper. All the lines are normalized in way to obtain the same energy flux of $0.9\times10^{-11}$~erg~cm$^{-2}$~s$^{-1}$ at 1~GeV. This is done only for comparison purposes and for all the cases the required energy in protons represent a small fraction of the explosion energy.} 
  
  \label{fig:breakenergy}
\end{figure}

First, to study the impact of the break energy $E_{\rm b}$ on the resulting gamma-ray spectrum, we fix spectral indices to $s_1 = 2.0$ (motivated by diffusive shock acceleration, DSA) and $s_2 = 2.75$ (motivated by observations and close to the value predicted by numerical simulations, \citealt{2020A&A...634A..59B}) and vary the break energy from 1 to 10 GeV. In Fig. \ref{fig:breakenergy} we illustrate how the shape of the expected gamma-ray spectrum changes depending on the value of the break energy. For each line $N_0$ is chosen in a way to obtain the same energy flux of gamma-rays of $0.9\times10^{-11}$~erg~cm$^{-2}$~s$^{-1}$ at 1~GeV. Modelled spectra clearly indicate that for $E_{\rm b}\gtrsim3$~GeV the expected turn-over shifts to higher energies significantly deviating from observations. In other words, if the particle acceleration at the shock produces a typical $-2$-power-law spectrum the current maximum energy to which particles can be accelerated cannot be larger than 3~GeV to reproduce the observed gamma-ray spectrum. This implies that the dynamical age of the remnant is very old and that it is far into post-adiabatic phase. The interaction with the dense material might enhance the escape of CRs further, besides the increase of the dynamical age. Neutral-charged collisions in the only partially ionized cloud might lead to an evanescence of the Alfv\'{e}n waves that confine the CRs \citep{Kulsrud.1971a}, lower the current maximum energy and modify the CR-spectrum \citep{Malkov.2013a}. This low break energy is also only marginally compatible with the lower limit implied from the observed radio spectrum (see Section~\ref{sec:naima}) requiring a significantly higher magnetic field.

\begin{figure}
  \centering
  \includegraphics[width=8.0cm]{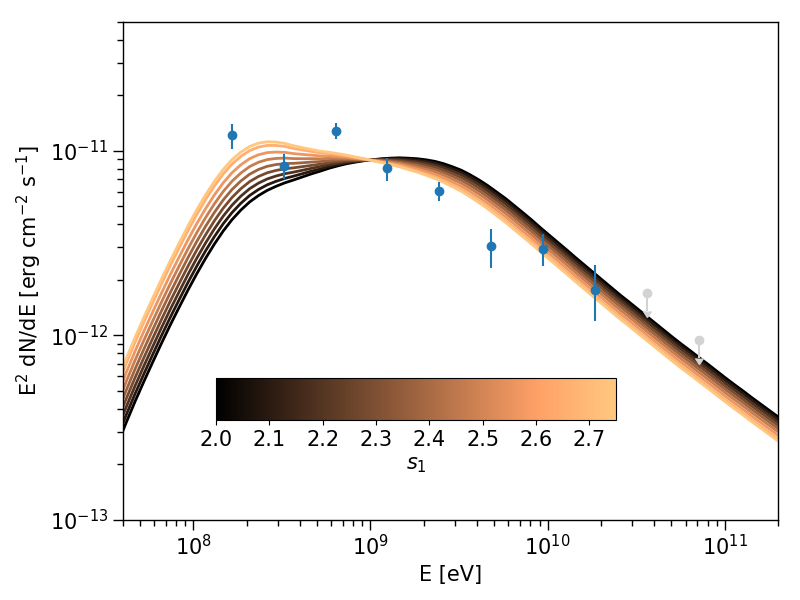}
  \caption{Same as Fig.~\ref{fig:breakenergy} but with fixed break energy at 10 GeV and varying $s_1$ from 2.0 to 2.75.}
  \label{fig:spectralindex}
\end{figure}

However, for old remnants the spectrum of accelerated particles may very well deviate from the featureless power-law with the spectral index $s=2$ obtained in DSA. It can become softer simply because the shock decelerates and becomes weaker resulting in a lower compression ratio. It can be also further modified as a result of substantial pressure of CRs which may provide feedback on the hydrodynamic evolution and modify the structure of the shock. In this case, the particle spectrum would have a concave shape with $s<2$ below a few GeV and $s>2$ above that \citep{2002APh....16..429B,2002ApJ...571..856M,1999ApJ...526..385B}. 
The soft spectrum of freshly accelerated particles might be another reason of such a low energy of the peak in the gamma-ray spectrum. In Fig. \ref{fig:spectralindex} we show how the shape of the gamma-ray spectrum changes for different values of $s_1$ spanning from $2.0$ to $2.75$. The break energy here is fixed at $E_{\rm b} = 10$~GeV. Each spectrum is normalized in the same way as in Fig.~\ref{fig:breakenergy}. Indeed, it can be seen that if the particle spectrum is soft enough the value of the break energy becomes less important. In fact, the observed gamma-ray spectrum can be relatively well explained with a single power-law with the spectral index of $2.75$. Soft spectrum, however, disagrees with radio observations which clearly imply a featureless power-law electron spectrum with $s = 1.8$. There is no obvious reason, why would the spectral shape of accelerated protons differ from the spectral shape of accelerated electrons, hence the proton spectrum in place is rather harder than $s=2.0$ than softer, implying an even lower break energy.

\begin{figure}
  \centering
  \includegraphics[width=8.0cm]{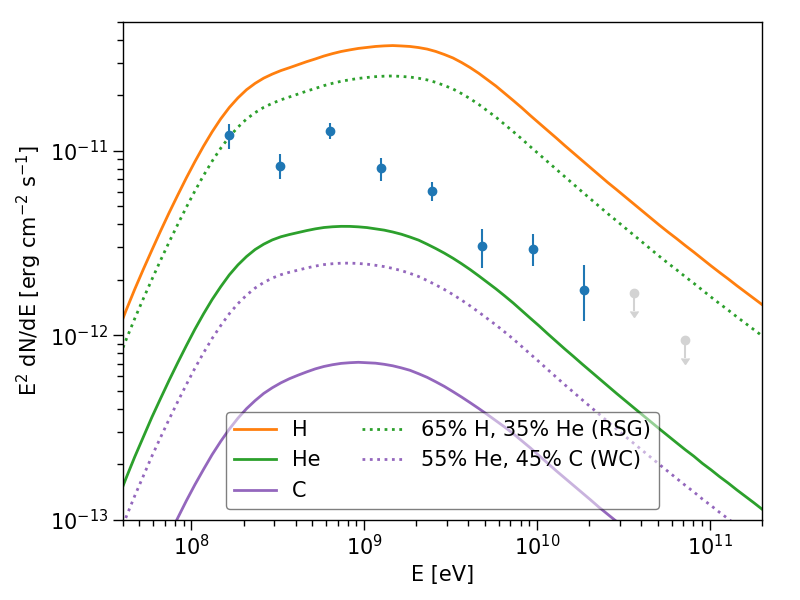}
  \caption{Same as Figs.~\ref{fig:breakenergy} and \ref{fig:spectralindex} but for different composition of accelerated particles for hadronic interactions. The solid lines show mono-elemental compositions for hydrogen (orange), helium (green) and carbon (purple) nuclei. Dotted lines represent mixed compositions (mass fractions) which reflect RSG and WC winds. The spectra are normalized to the same mass density of CRs (and hence the same injection efficiency) to demonstrate the difference in flux normalization when moving to heavier nuclei. The CR spectrum is assumed to follow a broken power-law in momentum with $s_1 = 2.0$, $s_2 = 2.75$ with a break at $Zp_{\rm b}$, where $p_{\rm b}$ defines the break momentum for the hydrogen spectrum corresponding the break energy of $10$~GeV. The target material is assumed to be of typical ISM composition with density of 400 cm$^{-3}$.}
  \label{fig:composition}
\end{figure}
\subsection{Heavy composition}
\label{sec:composition}
Above, we showed that both low break energy and soft spectrum of freshly accelerated particles can explain the shape of the observed gamma-ray spectrum with the characteristic turn-over at low energies. Both effects strongly suggest an old dynamical age of the remnant which is also in agreement with estimates of the density of the ambient medium which appears to be very high (see Sec. \ref{sec:co}). Nevertheless, the required break energy ($E_{\mathrm b} \lesssim 3$~GeV) is rather hard to get in time evolution models which predict a break energy around $10-100$~GeV and is only very marginally compatible with the featureless radio synchrotron spectrum up to 33\,GHz, while the soft spectral index disagrees with the observed radio spectrum which implies a much harder particle spectral index. 

To reconcile these discrepancies one might need to examine and take into account the elemental composition of the ambient medium in which the SNR is evolving. Indeed, heavy elements, both as CRs and as target material, result in a shift of the peak in the gamma-ray spectrum to lower energies \citep{2019ICRC...36..592P,2017PhRvD..95l3014B}. At the same time, heavy composition is expected for core-collapse SNRs which expand into the stellar wind of their progenitor stars, mostly red supergiants or Wolf-Rayet stars. Given that G39.2$-$0.3 is a core-collapse remnant of likely Type IIL/b with a RSG progenitor (see Section~\ref{sec:intro}), one can expect a fair fraction of heavy elements which can be accelerated at the shock and/or act as target material for hadronic interactions. In the following, we consider heavy nuclei only for CRs accelerated at the shock, keeping the target material of typical ISM composition with H to He ratio of 10:1. This is motivated by the idea that CRs residing in the SNR were accelerated at earlier stages of evolution while the SNR was still evolving inside the stellar wind bubble, while at this moment of time they are interacting with the dense cloud material of the ISM composition, which also suppresses farther particle acceleration due to a severe decrease of the shock velocity. We, however, note that abundance of the heavy elements in the target material would lead to the same effect boosting the modification of the gamma-ray spectrum.

To examine this we construct a momentum spectrum of accelerated particles in the form of:
\begin{equation}
    \frac{dN_i}{dp} = 
    \begin{cases}
        N_{0,i} p^{-s_1},& \text{if } p< Zp_{\rm b}\\
        N_{0,i} (Zp_{\rm b})^{-s_1+s_2} p^{-s_2},& \text{otherwise}.
    \end{cases}
    \label{eq:particle_spectrum2}
\end{equation}
where $i$ denotes the type of the particle and $Z$ is the charge number and $p_{\rm b}$ is the break momentum for hydrogen nuclei. The choice of a break momentum is based on the idea that it should reflect the maximum momentum to which particles can be accelerated at this moment of time \citep{2020A&A...634A..59B} and hence should scale with rigidity. We assume $s_1 = 2.0$, $s_2 = 2.75$ and $p_{\rm b}$ corresponds to the break kinetic energy of $E_{\mathrm b} = 10$~GeV for hydrogen nuclei. For helium this would correspond to the break at the kinetic energy of 18~GeV and for carbon at 55~GeV.

In Fig.~\ref{fig:composition} we show the resulting gamma-ray spectra for CRs consisting of only H (orange solid line), only He (green solid line), and only C (purple solid line), as well as for two mixed compositions which roughly correspond to the RSG \citep{2017A&A...605A..83D} and Wolf-Rayet \citep[][and references therein]{2019A&A...621A..92S} stellar winds. To calculate the expected gamma-ray spectrum we use the "pion decay" subroutine of the \textsc{RATPaC} code. 
In these calculations we assume that G39.2$-$0.3 is located at the distance of $6.2$~kpc and has a radius of $7$~pc which corresponds to its mean angular radio diameter of $7.8^\prime$ \citep{1990A&A...232..467P}. For all the cases we assume the density of the target material to be $400$~cm$^{-3}$ (as estimated for the molecular cloud) and keep the mass density of the CRs the same. The latter is done to illustrate how the normalization of the spectrum changes depending on the composition. The parameter space has, however, enough room to adjust the normalization of each model to correspond to the observed flux. The fraction of thermal particles injected into the acceleration process (above 10~MeV) adopted in the models shown in Fig.~\ref{fig:composition} is
\begin{equation}
    \eta = 4 \times 10^{-7} \frac{M_{\rm SWB}}{30\,M_\odot},
\end{equation}
where $M_{\rm SWB}$ is the total mass of the material in the wind bubble of the progenitor star. This corresponds to the injection parameter $\chi = 4.0$, which is defined as the multiple of the momentum at the peak of the thermal particle distribution, $p_{\mathrm{th}} = \sqrt{2mk_{\mathrm{B}}T}$, and determines the injection momentum of particles, $p_{\mathrm{inj}} = \chi p_{\mathrm{th}}$. 
This parameter can be as low as 3.5 \citep{2005MNRAS.361..907B} boosting the injection efficiency by two orders of magnitude, hence any of the illustrated models can effectively match the observations. The total energy in particles for the hydrogen alone case calculated above 10~MeV is $1.1\times10^{49}$ ergs and below that for other cases ($4.4\times10^{48}$ ergs for helium and $2.6\times10^{48}$ ergs for carbon), which leaves some room for maneuver also in terms of the energy budget. 

It can be seen that for heavy CRs the peak in the gamma-ray spectrum shifts to lower energies while the normalization decreases. Peak energies for hydrogen, helium and carbon are $1.5$~GeV, $0.7$~GeV, and $0.9$~GeV, respectively.
The RSG wind (green dotted lines in Fig. \ref{fig:composition}), which is still dominated by hydrogen, results in a similar spectrum to purely hydrogen CRs, with the peak energy staying essentially the same. On contrary, the carbon-loaded Wolf-Rayet wind (WC,  purple dotted lines in Fig. \ref{fig:composition})) significantly modifies the expected gamma-ray spectrum shifting the peak energy to $0.8$~GeV and much better reproducing the shape of the observed one. It should be noted here, that the total mass in the stellar wind bubble is dominated by the main-sequence stellar wind which has a usual ISM composition. This is, however, not necessarily a problem for this model, since the main-sequence wind should be crushed by the RSG and WR winds, pushing it to the edge of the bubble, and particles injected at the shock would originate from a heavier composition. Likewise the injection efficiency of particles can be increased not violating any physical constraints. 

Finally, the gamma-ray spectrum can be also boosted and modified by accounting for the particle acceleration at the reverse shock propagating into the heavy SNR ejecta. The contribution from the reverse shock is, however, expected to be negligible for the remnant of this size and age, because the density of the ejecta is expected to be rather low.

Taking into account the considerations above, a low-energy turn-over of the gamma-ray spectrum, which is observed in G39.2$-$0.3 and also in a few others SNRs, might be a tentative evidence that observed gamma-rays are produced in interactions of heavy nuclei. Moreover, precise measurements of gamma-ray spectra might in future serve to probe the composition of the media surrounding SNRs and even constrain the nature of their progenitor stars.

\subsection{Compression of Galactic CRs}

Another mechanism which might be responsible for the gamma-ray emission from SNRs interacting with dense clouds and is widely discussed in the literature in the context of the established hadronic emitters such as W~44 and IC~443 is compression and re-acceleration of Galactic CRs \citep{2010ApJ...723L.122U, 2015ApJ...806...71L, 2016A&A...595A..58C,2014ApJ...784L..35T, 2015ApJ...800..103T, 2019MNRAS.482.3843T}. The basic idea is that the interaction of the SNR shock with the dense cloud results in the formation of the radiative shell behind the shock front. The material behind the shock is adiabatically compressed to very high densities potentially boosting for pion-decay generated gamma-ray emission. The adiabatic compression of the pre-existing ambient Galactic CRs in the radiative shell enhances the CR spectrum both energizing particles and increasing the normalization of the spectrum. The compressed CR spectrum can be expressed then as \citep{2010ApJ...723L.122U}
\begin{equation}
    \label{eq:compression}
    n_\mathrm{comp}(p) = \xi^{2/3}n_\mathrm{GCR}(\xi^{-1/3}p)
\end{equation}
where $n_\mathrm{GCR}(p)$ is the density of Galactic CRs as a function of momentum and $\xi \equiv n_\mathrm{shell}/(rn_0)$ is the adiabatic compression ratio, with $n_\mathrm{shell}$ the density of the cooled gas in the shell, $n_0$ the density of the ambient medium (cloud), and $r$ the shock compression ratio. Additionally the pre-exisiting CRs can be further re-accelerated at the shock. This strong boost of the CR spectrum in combination with high gas density in the shell can result in substantial gamma-ray emission potentially explaining observed gamma-ray fluxes from aged SNRs interacting with dense clouds without necessity of direct particle acceleration at the shock. 

In the following we evaluate the feasibility of this scenario for the case of \snr. For simplicity we ignore re-acceleration of Galactic CRs taking into account only adiabatic compression, and also define the adiabatic compression ratio as the total compression between the radiative shell and the cloud, i.e. $\xi = n_\mathrm{shell}/n_0$. Then, assuming that the compression is limited by the magnetic pressure, with the magnetic field in the shell given by $B_\mathrm{shell} = \sqrt{2/3}\xi B_0$ (where $B_0$ is the magnetic field in the cloud) the compression ratio can be expressed as:
\begin{equation}
    \xi \simeq 94 \left[\frac{n_0}{1\,\mathrm{cm}^{-3}}\right]^{1/2}\left[\frac{B_0}{1\,\mu\mathrm{G}}\right]^{-1} \left[\frac{v_\mathrm{sh}}{10^7\,\mathrm{cm/s}}\right]
\end{equation}
For the proton CR spectrum we adopt the approximation of the observed proton flux proposed by \citet{2016Ap&SS.361...48B} imposed with a spectral hardening at higher energies \citep{2011Sci...332...69A, 2015PhRvL.114q1103A} as in \citealt{2016A&A...595A..58C}\footnote{Note that in \citet{2016A&A...595A..58C} some of the parameters for the proton and helium spectra adopted from \citet{2016Ap&SS.361...48B} are mistakenly swapped, which however does not change their results considerably.}:
\begin{align}
J_\mathrm{GCR}(E) = &3719 \frac{E^{1.03}}{\beta^2} \left(\frac{E^{1.21} + 0.77^{1.21}}{1+0.77^{1.21}}\right)^{-3.18} \\ \nonumber &\times\left[1+\left(\frac{E}{335}\right)^{\frac{0.119}{0.024}}\right]^{0.024}        
\end{align}
where $E$ is the kinetic energy of proton and $\beta$ is the proton velocity in $c$. The number density of CRs as a function of momentum is given then by
\begin{equation}
    n_\mathrm{GCR}(p) = \beta c \, n_\mathrm{GCR}(E) = 4\pi J_\mathrm{GCR}(E) 
\end{equation}
and the compressed spectrum can be found using Eq. \ref{eq:compression}.

\begin{figure}
  \centering
  \includegraphics[width=8.0cm]{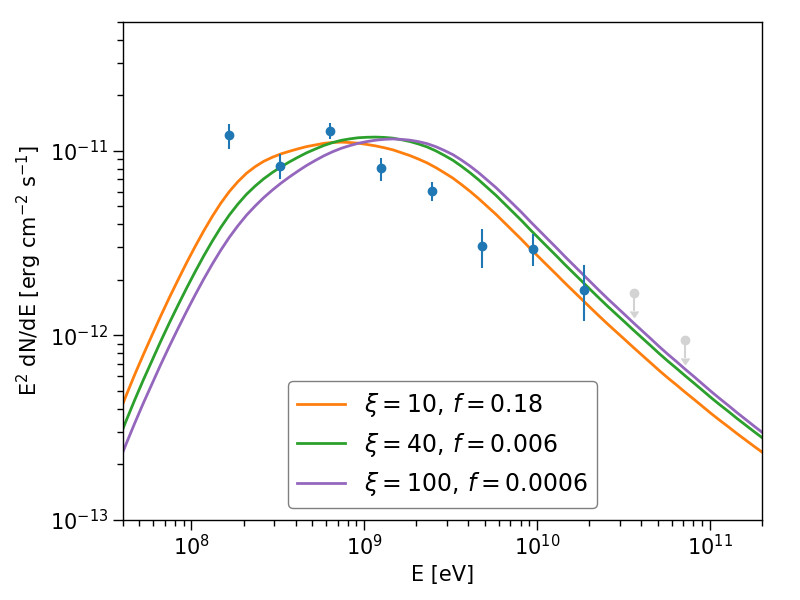}
  \caption{Spectral energy distribution of the modelled hadronic
gamma-ray emission generated by compressed Galactic CRs in the radiative shell for different values of the compression ratio. Simulated spectra are fit to the data by adjusting the filling factor $f$ which defines how much of the SNR volume is filled by
the radiative shell. Blue points represent $Fermi$-LAT data as in previous figures.}
  \label{fig:compression}  
\end{figure}

In Fig.~\ref{fig:compression} we show the computed gamma-ray emission from the compressed Galactic CRs for different values of $\xi$ and compare it to the observed spectrum. Simulated spectra are fit to the observational data by adjusting the volume filling factor $f$. The density in the cloud is assumed to be $n_0\sim400$~cm$^{-3}$ and the distance to the SNR and its radius are assumed to be $6.2$~kpc and $7$~pc respectively (same as in the previous section). It can be seen from the figure that for large $\xi$, the simulated gamma-ray spectra do not provide good fits for the data, peaking at larger energy ($\gtrsim 1$~GeV) than the observed spectrum. On the other hand, for lower $\xi$ we found that the filling factor needed to reach the level of the observed emission, is unrealistically high. Indeed, for $f = 0.18$ and $\xi = 10$ the total mass in the radiative shell is $\sim 4 \times 10^{4}\,M_\odot$. The total cloud mass that can possibly be accumulated within the volume of the SNR, $V_\mathrm{SNR}$, can be calculated as $V_\mathrm{SNR}n_0\mu_\mathrm{H} = 2\times10^4\,M_\odot$ ($\mu_\mathrm{H}$ is the mass per hydrogen for a typical ISM composition), which is half of the required mass for this scenario to work. A larger distance (and hence larger physical radius of the SNR) could partially alleviate these constraints, but in general this scenario seems rather unlikely to explain the emission observed from \snr.

Additionally, the above argument also applies for W~44. \citealt{2010ApJ...723L.122U} and \citealt{2016A&A...595A..58C} deducted a filling factor of $0.20$ and $0.15$ respectively, using a constant density of [5-10]$\times10^3$~cm$^{-3}$ behind the shock. For the assumed radius of the remnant of 12.5~pc, this corresponds to $\sim(2.5-4)\times10^5\,M_\odot$ which is almost an order of magnitude higher than the total cloud mass ($5.6\times10^4\,M_\odot$) for a cloud density of $200$~cm$^{-3}$, applied in both studies. These simple consideration suggests that even if adiabatic compression of Galactic CRs might significantly contribute to the gamma-ray emission, it most probably cannot alone explain the observed flux of W~44.

\section{Conclusions}
\label{sec:conclusions}

 We have identified \snr\ as a hadronic accelerator through its multi-wavelength properties. The combination of the radio and GeV data together with non-detection of non-thermal X-rays from the remnant clearly indicate that neither inverse Compton (due to a low amount of electrons) nor bremsstrahlung (due to the spectral shape) can be responsible for the observed GeV emission. On contrary, the hadronic scenario provides a relatively good fit to the data for a soft spectrum of protons with the spectral index of $\sim2.75$. Such a soft spectrum above some break energy which reflects the current maximum energy of accelerated protons is expected for dynamically old SNRs due to the escape of CRs and decrease of the acceleration efficiency. The total energy stored in CRs reaches $\sim 10^{49}$ erg, which reflects a few percent efficiency of converting kinetic energy into CRs, similarly to what has been observed in other SNR interacting with dense gas in large molecular clouds \citep{2016ApJS..224....8A}. \snr\ seems in general fainter than the typical interacting SNRs listed in the last $Fermi$ LAT SNR catalogue, when comparing with the radio emission (see Fig. 12 in \citealt{2016ApJS..224....8A}), but still within the main distribution.

The dense medium in which this SNR is evolving implies strong bremsstrahlung emission which in combination with observed GeV emission sets a lower limit on magnetic field of around $\sim100\mu$G. Such a strong magnetic field is not typical for evolved SNRs but expected for remnants interacting with molecular clouds as the thick medium compresses the shock region, resulting in an amplification of the magnetic field. High magnetic field in principle favours the acceleration of protons to high energies. However, the dense matter also slows down the shock, preventing the acceleration to go beyond a few tens of GeV. Even considering this effect, the low break energy required in \snr\ seems too low, when considering evolution models of SNRs and the featureless radio Synchrotron spectrum. We investigate further how to reproduce the observed spectral shape by considering several hypothesis. To explain the low-energy peak in the gamma-ray spectrum, the CR spectrum requires a very low break energy, i.e. current maximum energy of protons, of $E_{\mathrm b}\lesssim 3$~GeV  and/or a softer than typical DSA spectrum of protons from the acceleration process, $s_1\gtrsim2.3$. Both requirements point to the old dynamical age of the remnant, which means that it is already at the late stages of its evolution. This is not surprising given the interaction with the molecular cloud which drastically increases the density of the ambient medium. However, both of these requirements are not trivial to fulfill even for a dynamically old SNR.
 
On the other hand, the core-collapse nature of the SNR implies that heavier composition of the surrounding medium may be reflected in the resulting cosmic-ray and gamma-ray spectra. Hadronic interactions involving heavy nuclei result in a peak of the gamma-ray emission at significantly lower energies then proton-proton interactions. We show that accounting for the heavy composition of the circumstellar medium which is translated into the heavy composition of accelerated particles may help to explain the observed gamma-ray spectrum without need for unusually soft spectrum or low break energy, but requiring that the progenitor was a Wolf-Rayet star rather than a red supergiant. This result implies that precise measurements of gamma-ray emission from evolved core-collapse SNRs might potentially probe the composition of the surrounding environment and even the nature of the progenitor star. 

We also investigated a scenario in which pre-existing Galactic CRs are compressed within the radiative shell and emit gamma-rays through hadronic interactions. We found that an unrealistically large size of the shell is required to explain the observed gamma-ray emission, which imposes substantial difficulties of this model.
 
 \snr\ is one of the few (18) SNR detected using the GLIMPSE Legacy science program on the Spitzer Space Telescope \citep{2006AJ....131.1479R}, and therefore one of the brightest SNR in infrared wavelengths. The observed infrared radiation reveals a mixture of molecular and ionic shocks, which indicate a clear production of CRs in this type of SNRs. Indeed, the majority of the infrared-bright SNRs have a counterpart on the $Fermi$ LAT catalogue. Comparing the proton spectrum of this population of SNRs would help to understand CRs production in Type II SNRs.


\section*{Data Availability}
The $Fermi$ data underlying this article are available at https://fermi.gsfc.nasa.gov/ssc/data/access/lat/. The CO data were accessed from the MWISP survey and can be shared on reasonable request to the corresponding author.

\section*{Acknowledgements}
This research was supported by the Alexander von Humboldt Foundation. The authors would like to thank J. Li for useful discussions and to the anonymous referee for constructive comments.



\bibliographystyle{mnras}
\bibliography{snrg39}

\begin{thebibliography}{}
\makeatletter
\relax
\def\mn@urlcharsother{\let\do\@makeother \do\$\do\&\do\#\do\^\do\_\do\%\do\~}
\def\mn@doi{\begingroup\mn@urlcharsother \@ifnextchar [ {\mn@doi@}
  {\mn@doi@[]}}
\def\mn@doi@[#1]#2{\def\@tempa{#1}\ifx\@tempa\@empty \href
  {http://dx.doi.org/#2} {doi:#2}\else \href {http://dx.doi.org/#2} {#1}\fi
  \endgroup}
\def\mn@eprint#1#2{\mn@eprint@#1:#2::\@nil}
\def\mn@eprint@arXiv#1{\href {http://arxiv.org/abs/#1} {{\tt arXiv:#1}}}
\def\mn@eprint@dblp#1{\href {http://dblp.uni-trier.de/rec/bibtex/#1.xml}
  {dblp:#1}}
\def\mn@eprint@#1:#2:#3:#4\@nil{\def\@tempa {#1}\def\@tempb {#2}\def\@tempc
  {#3}\ifx \@tempc \@empty \let \@tempc \@tempb \let \@tempb \@tempa \fi \ifx
  \@tempb \@empty \def\@tempb {arXiv}\fi \@ifundefined
  {mn@eprint@\@tempb}{\@tempb:\@tempc}{\expandafter \expandafter \csname
  mn@eprint@\@tempb\endcsname \expandafter{\@tempc}}}

\bibitem[\protect\citeauthoryear{{Abdo} et~al.,}{{Abdo}
  et~al.}{2011}]{2011ApJ...734...28A}
{Abdo} A.~A.,  et~al., 2011, \mn@doi [\apj] {10.1088/0004-637X/734/1/28}, \href
  {https://ui.adsabs.harvard.edu/abs/2011ApJ...734...28A} {734, 28}

\bibitem[\protect\citeauthoryear{{Acero} et~al.,}{{Acero}
  et~al.}{2016}]{2016ApJS..224....8A}
{Acero} F.,  et~al., 2016, \mn@doi [\apjs] {10.3847/0067-0049/224/1/8}, \href
  {https://ui.adsabs.harvard.edu/abs/2016ApJS..224....8A} {224, 8}

\bibitem[\protect\citeauthoryear{{Ackermann} et~al.,}{{Ackermann}
  et~al.}{2013}]{2013Sci...339..807A}
{Ackermann} M.,  et~al., 2013, \mn@doi [Science] {10.1126/science.1231160},
  \href {https://ui.adsabs.harvard.edu/abs/2013Sci...339..807A} {339, 807}

\bibitem[\protect\citeauthoryear{{Adriani} et~al.,}{{Adriani}
  et~al.}{2011}]{2011Sci...332...69A}
{Adriani} O.,  et~al., 2011, \mn@doi [Science] {10.1126/science.1199172}, \href
  {https://ui.adsabs.harvard.edu/abs/2011Sci...332...69A} {332, 69}

\bibitem[\protect\citeauthoryear{{Aguilar} et~al.,}{{Aguilar}
  et~al.}{2015}]{2015PhRvL.114q1103A}
{Aguilar} M.,  et~al., 2015, \mn@doi [\prl] {10.1103/PhysRevLett.114.171103},
  \href {https://ui.adsabs.harvard.edu/abs/2015PhRvL.114q1103A} {114, 171103}

\bibitem[\protect\citeauthoryear{{Aharonian} \& {Atoyan}}{{Aharonian} \&
  {Atoyan}}{1996}]{1996A&A...309..917A}
{Aharonian} F.~A.,  {Atoyan} A.~M.,  1996, \aap, \href
  {https://ui.adsabs.harvard.edu/abs/1996A&A...309..917A} {309, 917}

\bibitem[\protect\citeauthoryear{{Ahnen} et~al.,}{{Ahnen}
  et~al.}{2017}]{2017MNRAS.472.2956A}
{Ahnen} M.~L.,  et~al., 2017, \mn@doi [\mnras] {10.1093/mnras/stx2079}, \href
  {https://ui.adsabs.harvard.edu/abs/2017MNRAS.472.2956A} {472, 2956}

\bibitem[\protect\citeauthoryear{{Ambrogi}, {Zanin}, {Casanova}, {De O{\~n}a
  Wilhelmi}, {Peron}  \& {Aharonian}}{{Ambrogi}
  et~al.}{2019}]{2019A&A...623A..86A}
{Ambrogi} L.,  {Zanin} R.,  {Casanova} S.,  {De O{\~n}a Wilhelmi} E.,  {Peron}
  G.,   {Aharonian} F.,  2019, \mn@doi [\aap] {10.1051/0004-6361/201833985},
  \href {https://ui.adsabs.harvard.edu/abs/2019A&A...623A..86A} {623, A86}

\bibitem[\protect\citeauthoryear{{Anderson} \& {Rudnick}}{{Anderson} \&
  {Rudnick}}{1993}]{1993ApJ...408..514A}
{Anderson} M.~C.,  {Rudnick} L.,  1993, \mn@doi [\apj] {10.1086/172609}, \href
  {https://ui.adsabs.harvard.edu/abs/1993ApJ...408..514A} {408, 514}

\bibitem[\protect\citeauthoryear{{Archambault} et~al.,}{{Archambault}
  et~al.}{2017}]{2017ApJ...836...23A}
{Archambault} S.,  et~al., 2017, \mn@doi [\apj] {10.3847/1538-4357/836/1/23},
  \href {https://ui.adsabs.harvard.edu/abs/2017ApJ...836...23A} {836, 23}

\bibitem[\protect\citeauthoryear{{Banik} \& {Bhadra}}{{Banik} \&
  {Bhadra}}{2017}]{2017PhRvD..95l3014B}
{Banik} P.,  {Bhadra} A.,  2017, \mn@doi [\prd] {10.1103/PhysRevD.95.123014},
  \href {https://ui.adsabs.harvard.edu/abs/2017PhRvD..95l3014B} {95, 123014}

\bibitem[\protect\citeauthoryear{{Bass} et~al.,}{{Bass}
  et~al.}{1998}]{1998PrPNP..41..255B}
{Bass} S.~A.,  et~al., 1998, \mn@doi [Progress in Particle and Nuclear Physics]
  {10.1016/S0146-6410(98)00058-1}, \href
  {https://ui.adsabs.harvard.edu/abs/1998PrPNP..41..255B} {41, 255}

\bibitem[\protect\citeauthoryear{{Becker} \& {Kundu}}{{Becker} \&
  {Kundu}}{1975}]{1975AJ.....80..679B}
{Becker} R.~H.,  {Kundu} M.~R.,  1975, \mn@doi [\aj] {10.1086/111797}, \href
  {https://ui.adsabs.harvard.edu/abs/1975AJ.....80..679B} {80, 679}

\bibitem[\protect\citeauthoryear{{Berezhko} \& {Ellison}}{{Berezhko} \&
  {Ellison}}{1999}]{1999ApJ...526..385B}
{Berezhko} E.~G.,  {Ellison} D.~C.,  1999, \mn@doi [\apj] {10.1086/307993},
  \href {https://ui.adsabs.harvard.edu/abs/1999ApJ...526..385B} {526, 385}

\bibitem[\protect\citeauthoryear{{Bhatt}, {Sushch}, {Pohl}, {Fedynitch}, {Das},
  {Brose}, {Plotko}  \& {Meyer}}{{Bhatt} et~al.}{2020}]{2019ICRC...36..592P}
{Bhatt} M.,  {Sushch} I.,  {Pohl} M.,  {Fedynitch} A.,  {Das} S.,  {Brose} R.,
  {Plotko} P.,   {Meyer} D. M.~A.,  2020, \mn@doi [Astroparticle Physics]
  {10.1016/j.astropartphys.2020.102490}, \href
  {https://ui.adsabs.harvard.edu/abs/2020APh...12302490B} {123, 102490}

\bibitem[\protect\citeauthoryear{{Bisschoff} \& {Potgieter}}{{Bisschoff} \&
  {Potgieter}}{2016}]{2016Ap&SS.361...48B}
{Bisschoff} D.,  {Potgieter} M.~S.,  2016, \mn@doi [\apss]
  {10.1007/s10509-015-2633-8}, \href
  {https://ui.adsabs.harvard.edu/abs/2016Ap&SS.361...48B} {361, 48}

\bibitem[\protect\citeauthoryear{{Blasi}}{{Blasi}}{2002}]{2002APh....16..429B}
{Blasi} P.,  2002, \mn@doi [Astroparticle Physics]
  {10.1016/S0927-6505(01)00127-X}, \href
  {http://adsabs.harvard.edu/abs/2002APh....16..429B} {16, 429}

\bibitem[\protect\citeauthoryear{{Blasi}, {Gabici}  \& {Vannoni}}{{Blasi}
  et~al.}{2005}]{2005MNRAS.361..907B}
{Blasi} P.,  {Gabici} S.,   {Vannoni} G.,  2005, \mn@doi [\mnras]
  {10.1111/j.1365-2966.2005.09227.x}, \href
  {https://ui.adsabs.harvard.edu/abs/2005MNRAS.361..907B} {361, 907}

\bibitem[\protect\citeauthoryear{Bleicher et~al.,}{Bleicher
  et~al.}{1999}]{bleicher1999relativistic}
Bleicher M.,  et~al., 1999, Journal of Physics G: Nuclear and Particle Physics,
  25, 1859

\bibitem[\protect\citeauthoryear{{Bolatto}, {Wolfire}  \& {Leroy}}{{Bolatto}
  et~al.}{2013}]{2013ARA&A..51..207B}
{Bolatto} A.~D.,  {Wolfire} M.,   {Leroy} A.~K.,  2013, \mn@doi [\araa]
  {10.1146/annurev-astro-082812-140944}, \href
  {https://ui.adsabs.harvard.edu/abs/2013ARA&A..51..207B} {51, 207}

\bibitem[\protect\citeauthoryear{{Brose}, {Telezhinsky}  \& {Pohl}}{{Brose}
  et~al.}{2016}]{2016A&A...593A..20B}
{Brose} R.,  {Telezhinsky} I.,   {Pohl} M.,  2016, \mn@doi [\aap]
  {10.1051/0004-6361/201527345}, \href
  {http://adsabs.harvard.edu/abs/2016A%26A...593A..20B} {593, A20}

\bibitem[\protect\citeauthoryear{{Brose}, {Pohl}, {Sushch}, {Petruk}  \&
  {Kuzyo}}{{Brose} et~al.}{2020}]{2020A&A...634A..59B}
{Brose} R.,  {Pohl} M.,  {Sushch} I.,  {Petruk} O.,   {Kuzyo} T.,  2020,
  \mn@doi [\aap] {10.1051/0004-6361/201936567}, \href
  {https://ui.adsabs.harvard.edu/abs/2020A&A...634A..59B} {634, A59}

\bibitem[\protect\citeauthoryear{{Cardillo}, {Amato}  \& {Blasi}}{{Cardillo}
  et~al.}{2016}]{2016A&A...595A..58C}
{Cardillo} M.,  {Amato} E.,   {Blasi} P.,  2016, \mn@doi [\aap]
  {10.1051/0004-6361/201628669}, \href
  {https://ui.adsabs.harvard.edu/abs/2016A&A...595A..58C} {595, A58}

\bibitem[\protect\citeauthoryear{{Caswell}, {Murray}, {Roger}, {Cole}  \&
  {Cooke}}{{Caswell} et~al.}{1975}]{1975A&A....45..239C}
{Caswell} J.~L.,  {Murray} J.~D.,  {Roger} R.~S.,  {Cole} D.~J.,   {Cooke}
  D.~J.,  1975, \aap, \href
  {https://ui.adsabs.harvard.edu/abs/1975A&A....45..239C} {45, 239}

\bibitem[\protect\citeauthoryear{{Celli}, {Morlino}, {Gabici}  \&
  {Aharonian}}{{Celli} et~al.}{2019a}]{2019MNRAS.487.3199C}
{Celli} S.,  {Morlino} G.,  {Gabici} S.,   {Aharonian} F.~A.,  2019a, \mn@doi
  [\mnras] {10.1093/mnras/stz1425}, \href
  {https://ui.adsabs.harvard.edu/abs/2019MNRAS.487.3199C} {487, 3199}

\bibitem[\protect\citeauthoryear{{Celli}, {Morlino}, {Gabici}  \&
  {Aharonian}}{{Celli} et~al.}{2019b}]{2019MNRAS.490.4317C}
{Celli} S.,  {Morlino} G.,  {Gabici} S.,   {Aharonian} F.~A.,  2019b, \mn@doi
  [\mnras] {10.1093/mnras/stz2897}, \href
  {https://ui.adsabs.harvard.edu/abs/2019MNRAS.490.4317C} {490, 4317}

\bibitem[\protect\citeauthoryear{{Cruciani} et~al.,}{{Cruciani}
  et~al.}{2016}]{2016MNRAS.459.4224C}
{Cruciani} A.,  et~al., 2016, \mn@doi [\mnras] {10.1093/mnras/stw839}, \href
  {https://ui.adsabs.harvard.edu/abs/2016MNRAS.459.4224C} {459, 4224}

\bibitem[\protect\citeauthoryear{{Crutcher}, {Wandelt}, {Heiles}, {Falgarone}
  \& {Troland}}{{Crutcher} et~al.}{2010}]{2010ApJ...725..466C}
{Crutcher} R.~M.,  {Wandelt} B.,  {Heiles} C.,  {Falgarone} E.,   {Troland}
  T.~H.,  2010, \mn@doi [\apj] {10.1088/0004-637X/725/1/466}, \href
  {https://ui.adsabs.harvard.edu/abs/2010ApJ...725..466C} {725, 466}

\bibitem[\protect\citeauthoryear{{Dame}, {Hartmann}  \& {Thaddeus}}{{Dame}
  et~al.}{2001}]{2001ApJ...547..792D}
{Dame} T.~M.,  {Hartmann} D.,   {Thaddeus} P.,  2001, \mn@doi [\apj]
  {10.1086/318388}, \href
  {https://ui.adsabs.harvard.edu/abs/2001ApJ...547..792D} {547, 792}

\bibitem[\protect\citeauthoryear{{Dessart}, {John Hillier}  \&
  {Audit}}{{Dessart} et~al.}{2017}]{2017A&A...605A..83D}
{Dessart} L.,  {John Hillier} D.,   {Audit} E.,  2017, \mn@doi [\aap]
  {10.1051/0004-6361/201730942}, \href
  {https://ui.adsabs.harvard.edu/abs/2017A&A...605A..83D} {605, A83}

\bibitem[\protect\citeauthoryear{{Ferrand} \& {Safi-Harb}}{{Ferrand} \&
  {Safi-Harb}}{2012}]{2012AdSpR..49.1313F}
{Ferrand} G.,  {Safi-Harb} S.,  2012, \mn@doi [Advances in Space Research]
  {10.1016/j.asr.2012.02.004}, \href
  {https://ui.adsabs.harvard.edu/abs/2012AdSpR..49.1313F} {49, 1313}

\bibitem[\protect\citeauthoryear{{Fraschetti}}{{Fraschetti}}{2014}]{2014NIMPA.742..169F}
{Fraschetti} F.,  2014, \mn@doi [Nuclear Instruments and Methods in Physics
  Research A] {10.1016/j.nima.2013.11.066}, \href
  {http://adsabs.harvard.edu/abs/2014NIMPA.742..169F} {742, 169}

\bibitem[\protect\citeauthoryear{{Gabici}, {Gaggero}  \& {Zandanel}}{{Gabici}
  et~al.}{2016}]{2016arXiv161007638G}
{Gabici} S.,  {Gaggero} D.,   {Zandanel} F.,  2016, arXiv e-prints, \href
  {https://ui.adsabs.harvard.edu/abs/2016arXiv161007638G} {p. arXiv:1610.07638}

\bibitem[\protect\citeauthoryear{{Giacalone} \& {Jokipii}}{{Giacalone} \&
  {Jokipii}}{2007}]{2007ApJ...663L..41G}
{Giacalone} J.,  {Jokipii} J.~R.,  2007, \mn@doi [\apjl] {10.1086/519994},
  \href {https://ui.adsabs.harvard.edu/abs/2007ApJ...663L..41G} {663, L41}

\bibitem[\protect\citeauthoryear{{Giuliani} \& {AGILE Team}}{{Giuliani} \&
  {AGILE Team}}{2011}]{2011MmSAI..82..747G}
{Giuliani} G.,  {AGILE Team} 2011, \memsai, \href
  {https://ui.adsabs.harvard.edu/abs/2011MmSAI..82..747G} {82, 747}

\bibitem[\protect\citeauthoryear{{Green}, {Frail}, {Goss}  \&
  {Otrupcek}}{{Green} et~al.}{1997}]{1997AJ....114.2058G}
{Green} A.~J.,  {Frail} D.~A.,  {Goss} W.~M.,   {Otrupcek} R.,  1997, \mn@doi
  [\aj] {10.1086/118626}, \href
  {https://ui.adsabs.harvard.edu/abs/1997AJ....114.2058G} {114, 2058}

\bibitem[\protect\citeauthoryear{{H.~E.~S.~S. Collaboration}
  et~al.,}{{H.~E.~S.~S. Collaboration} et~al.}{2018}]{2018A&A...612A...3H}
{H.~E.~S.~S. Collaboration} et~al., 2018, \mn@doi [\aap]
  {10.1051/0004-6361/201732125}, \href
  {https://ui.adsabs.harvard.edu/abs/2018A&A...612A...3H} {612, A3}

\bibitem[\protect\citeauthoryear{{Harrus} \& {Slane}}{{Harrus} \&
  {Slane}}{1999}]{1999ApJ...516..811H}
{Harrus} I.~M.,  {Slane} P.~O.,  1999, \mn@doi [\apj] {10.1086/307138}, \href
  {https://ui.adsabs.harvard.edu/abs/1999ApJ...516..811H} {516, 811}

\bibitem[\protect\citeauthoryear{{Jogler} \& {Funk}}{{Jogler} \&
  {Funk}}{2016}]{2016ApJ...816..100J}
{Jogler} T.,  {Funk} S.,  2016, \mn@doi [\apj] {10.3847/0004-637X/816/2/100},
  \href {https://ui.adsabs.harvard.edu/abs/2016ApJ...816..100J} {816, 100}

\bibitem[\protect\citeauthoryear{{Kulsrud} \& {Cesarsky}}{{Kulsrud} \&
  {Cesarsky}}{1971}]{Kulsrud.1971a}
{Kulsrud} R.~M.,  {Cesarsky} C.~J.,  1971, aplett, \href
  {http://adsabs.harvard.edu/abs/1971ApL.....8..189K} {8, 189}

\bibitem[\protect\citeauthoryear{{Lee}, {Moon}, {Koo}, {Lee}  \&
  {Matthews}}{{Lee} et~al.}{2009}]{2009ApJ...691.1042L}
{Lee} H.-G.,  {Moon} D.-S.,  {Koo} B.-C.,  {Lee} J.-J.,   {Matthews} K.,  2009,
  \mn@doi [\apj] {10.1088/0004-637X/691/2/1042}, \href
  {https://ui.adsabs.harvard.edu/abs/2009ApJ...691.1042L} {691, 1042}

\bibitem[\protect\citeauthoryear{{Lee}, {Patnaude}, {Raymond}, {Nagataki},
  {Slane}  \& {Ellison}}{{Lee} et~al.}{2015}]{2015ApJ...806...71L}
{Lee} S.-H.,  {Patnaude} D.~J.,  {Raymond} J.~C.,  {Nagataki} S.,  {Slane}
  P.~O.,   {Ellison} D.~C.,  2015, \mn@doi [\apj] {10.1088/0004-637X/806/1/71},
  \href {https://ui.adsabs.harvard.edu/abs/2015ApJ...806...71L} {806, 71}

\bibitem[\protect\citeauthoryear{{Malkov}, {Diamond}  \& {Jones}}{{Malkov}
  et~al.}{2002}]{2002ApJ...571..856M}
{Malkov} M.~A.,  {Diamond} P.~H.,   {Jones} T.~W.,  2002, \mn@doi [\apj]
  {10.1086/340097}, \href
  {https://ui.adsabs.harvard.edu/abs/2002ApJ...571..856M} {571, 856}

\bibitem[\protect\citeauthoryear{{Malkov}, {Diamond}  \& {Sagdeev}}{{Malkov}
  et~al.}{2011}]{2011NatCo...2..194M}
{Malkov} M.~A.,  {Diamond} P.~H.,   {Sagdeev} R.~Z.,  2011, \mn@doi [Nature
  Communications] {10.1038/ncomms1195}, \href
  {https://ui.adsabs.harvard.edu/abs/2011NatCo...2..194M} {2, 194}

\bibitem[\protect\citeauthoryear{{Malkov}, {Diamond}, {Sagdeev}, {Aharonian}
  \& {Moskalenko}}{{Malkov} et~al.}{2013}]{Malkov.2013a}
{Malkov} M.~A.,  {Diamond} P.~H.,  {Sagdeev} R.~Z.,  {Aharonian} F.~A.,
  {Moskalenko} I.~V.,  2013, \mn@doi [apj] {10.1088/0004-637X/768/1/73}, \href
  {http://adsabs.harvard.edu/abs/2013ApJ...768...73M} {768, 73}

\bibitem[\protect\citeauthoryear{{Morlino} \& {Caprioli}}{{Morlino} \&
  {Caprioli}}{2012}]{2012A&A...538A..81M}
{Morlino} G.,  {Caprioli} D.,  2012, \mn@doi [\aap]
  {10.1051/0004-6361/201117855}, \href
  {https://ui.adsabs.harvard.edu/abs/2012A&A...538A..81M} {538, A81}

\bibitem[\protect\citeauthoryear{{Ohira}, {Murase}  \& {Yamazaki}}{{Ohira}
  et~al.}{2011}]{2011MNRAS.410.1577O}
{Ohira} Y.,  {Murase} K.,   {Yamazaki} R.,  2011, \mn@doi [\mnras]
  {10.1111/j.1365-2966.2010.17539.x}, \href
  {https://ui.adsabs.harvard.edu/abs/2011MNRAS.410.1577O} {410, 1577}

\bibitem[\protect\citeauthoryear{{Olbert}, {Keohane}, {Arnaud}, {Dyer},
  {Reynolds}  \& {Safi-Harb}}{{Olbert} et~al.}{2003}]{2003ApJ...592L..45O}
{Olbert} C.~M.,  {Keohane} J.~W.,  {Arnaud} K.~A.,  {Dyer} K.~K.,  {Reynolds}
  S.~P.,   {Safi-Harb} S.,  2003, \mn@doi [\apjl] {10.1086/377348}, \href
  {https://ui.adsabs.harvard.edu/abs/2003ApJ...592L..45O} {592, L45}

\bibitem[\protect\citeauthoryear{{Patnaik}, {Hunt}, {Salter}, {Shaver}  \&
  {Velusamy}}{{Patnaik} et~al.}{1990}]{1990A&A...232..467P}
{Patnaik} A.~R.,  {Hunt} G.~C.,  {Salter} C.~J.,  {Shaver} P.~A.,   {Velusamy}
  T.,  1990, \aap, \href
  {https://ui.adsabs.harvard.edu/abs/1990A&A...232..467P} {232, 467}

\bibitem[\protect\citeauthoryear{{Reach} et~al.,}{{Reach}
  et~al.}{2006}]{2006AJ....131.1479R}
{Reach} W.~T.,  et~al., 2006, \mn@doi [\aj] {10.1086/499306}, \href
  {https://ui.adsabs.harvard.edu/abs/2006AJ....131.1479R} {131, 1479}

\bibitem[\protect\citeauthoryear{Roesler, Engel  \& Ranft}{Roesler
  et~al.}{2001}]{roesler2001monte}
Roesler S.,  Engel R.,   Ranft J.,  2001, in , Advanced Monte Carlo for
  radiation physics, particle transport simulation and applications.
Springer, pp 1033--1038

\bibitem[\protect\citeauthoryear{{Sander}, {Hamann}, {Todt}, {Hainich},
  {Shenar}, {Ramachandran}  \& {Oskinova}}{{Sander}
  et~al.}{2019}]{2019A&A...621A..92S}
{Sander} A.~A.~C.,  {Hamann} W.~R.,  {Todt} H.,  {Hainich} R.,  {Shenar} T.,
  {Ramachandran} V.,   {Oskinova} L.~M.,  2019, \mn@doi [\aap]
  {10.1051/0004-6361/201833712}, \href
  {https://ui.adsabs.harvard.edu/abs/2019A&A...621A..92S} {621, A92}

\bibitem[\protect\citeauthoryear{{Sezer}, {Ergin}, {Cesur}, {Tanaka}, {Kisaka},
  {Ohira}  \& {Yamazaki}}{{Sezer} et~al.}{2020}]{2020MNRAS.492.1484S}
{Sezer} A.,  {Ergin} T.,  {Cesur} N.,  {Tanaka} S.~J.,  {Kisaka} S.,  {Ohira}
  Y.,   {Yamazaki} R.,  2020, \mn@doi [\mnras] {10.1093/mnras/stz3571}, \href
  {https://ui.adsabs.harvard.edu/abs/2020MNRAS.492.1484S} {492, 1484}

\bibitem[\protect\citeauthoryear{{Su}, {Chen}, {Yang}, {Koo}, {Zhou}, {Lu},
  {Jeong}  \& {DeLaney}}{{Su} et~al.}{2011}]{2011ApJ...727...43S}
{Su} Y.,  {Chen} Y.,  {Yang} J.,  {Koo} B.-C.,  {Zhou} X.,  {Lu} D.-R.,
  {Jeong} I.-G.,   {DeLaney} T.,  2011, \mn@doi [\apj]
  {10.1088/0004-637X/727/1/43}, \href
  {https://ui.adsabs.harvard.edu/abs/2011ApJ...727...43S} {727, 43}

\bibitem[\protect\citeauthoryear{{Su} et~al.,}{{Su}
  et~al.}{2019}]{2019ApJS..240....9S}
{Su} Y.,  et~al., 2019, \mn@doi [\apjs] {10.3847/1538-4365/aaf1c8}, \href
  {https://ui.adsabs.harvard.edu/abs/2019ApJS..240....9S} {240, 9}

\bibitem[\protect\citeauthoryear{{Sushch}, {Brose}  \& {Pohl}}{{Sushch}
  et~al.}{2018}]{2018A&A...618A.155S}
{Sushch} I.,  {Brose} R.,   {Pohl} M.,  2018, \mn@doi [\aap]
  {10.1051/0004-6361/201832879}, \href
  {https://ui.adsabs.harvard.edu/abs/2018A&A...618A.155S} {618, A155}

\bibitem[\protect\citeauthoryear{{Tang}}{{Tang}}{2019}]{2019MNRAS.482.3843T}
{Tang} X.,  2019, \mn@doi [\mnras] {10.1093/mnras/sty1042}, \href
  {https://ui.adsabs.harvard.edu/abs/2019MNRAS.482.3843T} {482, 3843}

\bibitem[\protect\citeauthoryear{{Tang} \& {Chevalier}}{{Tang} \&
  {Chevalier}}{2014}]{2014ApJ...784L..35T}
{Tang} X.,  {Chevalier} R.~A.,  2014, \mn@doi [\apjl]
  {10.1088/2041-8205/784/2/L35}, \href
  {https://ui.adsabs.harvard.edu/abs/2014ApJ...784L..35T} {784, L35}

\bibitem[\protect\citeauthoryear{{Tang} \& {Chevalier}}{{Tang} \&
  {Chevalier}}{2015}]{2015ApJ...800..103T}
{Tang} X.,  {Chevalier} R.~A.,  2015, \mn@doi [\apj]
  {10.1088/0004-637X/800/2/103}, \href
  {https://ui.adsabs.harvard.edu/abs/2015ApJ...800..103T} {800, 103}

\bibitem[\protect\citeauthoryear{{Telezhinsky}, {Dwarkadas}  \&
  {Pohl}}{{Telezhinsky} et~al.}{2012}]{2012APh....35..300T}
{Telezhinsky} I.,  {Dwarkadas} V.~V.,   {Pohl} M.,  2012, \mn@doi
  [Astroparticle Physics] {10.1016/j.astropartphys.2011.10.001}, \href
  {http://adsabs.harvard.edu/abs/2012APh....35..300T} {35, 300}

\bibitem[\protect\citeauthoryear{{Telezhinsky}, {Dwarkadas}  \&
  {Pohl}}{{Telezhinsky} et~al.}{2013}]{2013A&A...552A.102T}
{Telezhinsky} I.,  {Dwarkadas} V.~V.,   {Pohl} M.,  2013, \mn@doi [\aap]
  {10.1051/0004-6361/201220740}, \href
  {http://adsabs.harvard.edu/abs/2013A%26A...552A.102T} {552, A102}

\bibitem[\protect\citeauthoryear{{The Fermi-LAT collaboration}}{{The Fermi-LAT
  collaboration}}{2019}]{2019arXiv190210045T}
{The Fermi-LAT collaboration} 2019, arXiv e-prints, \href
  {https://ui.adsabs.harvard.edu/abs/2019arXiv190210045T} {p. arXiv:1902.10045}

\bibitem[\protect\citeauthoryear{{Uchiyama}, {Blandford}, {Funk}, {Tajima}  \&
  {Tanaka}}{{Uchiyama} et~al.}{2010}]{2010ApJ...723L.122U}
{Uchiyama} Y.,  {Blandford} R.~D.,  {Funk} S.,  {Tajima} H.,   {Tanaka} T.,
  2010, \mn@doi [\apjl] {10.1088/2041-8205/723/1/L122}, \href
  {https://ui.adsabs.harvard.edu/abs/2010ApJ...723L.122U} {723, L122}

\bibitem[\protect\citeauthoryear{{Zabalza}}{{Zabalza}}{2017}]{2017ascl.soft08022Z}
{Zabalza} V.,  2017, {Naima: Derivation of non-thermal particle distributions
  through MCMC spectral fitting} (\mn@eprint {ascl} {1708.022})

\bibitem[\protect\citeauthoryear{{Zeng}, {Xin}  \& {Liu}}{{Zeng}
  et~al.}{2019}]{2019ApJ...874...50Z}
{Zeng} H.,  {Xin} Y.,   {Liu} S.,  2019, \mn@doi [\apj]
  {10.3847/1538-4357/aaf392}, \href
  {https://ui.adsabs.harvard.edu/abs/2019ApJ...874...50Z} {874, 50}

\bibitem[\protect\citeauthoryear{{Zirakashvili}, {Aharonian}, {Yang},
  {O{\~n}a-Wilhelmi}  \& {Tuffs}}{{Zirakashvili}
  et~al.}{2014}]{2014ApJ...785..130Z}
{Zirakashvili} V.~N.,  {Aharonian} F.~A.,  {Yang} R.,  {O{\~n}a-Wilhelmi} E.,
  {Tuffs} R.~J.,  2014, \mn@doi [\apj] {10.1088/0004-637X/785/2/130}, \href
  {https://ui.adsabs.harvard.edu/abs/2014ApJ...785..130Z} {785, 130}

\makeatother
\end{thebibliography}



\bsp	
\label{lastpage}
\end{document}